# AI Alignment vs. AI Ethical Treatment: 10 Challenges

Adam Bradley (Department of Philosophy and
Hong Kong Catastrophic Risk Centre, Lingnan University)
&
Bradford Saad (Global Priorities Institute, University of Oxford)[1]

**Abstract**: A morally acceptable course of AI development should avoid two dangers: creating unaligned AI systems that pose a threat to humanity and mistreating AI systems that merit moral consideration in their own right. This paper argues these two dangers interact and that if we create AI systems that merit moral consideration, simultaneously avoiding both of these dangers would be extremely challenging. While our argument is straightforward and supported by a wide range of pretheoretical moral judgments, it has far-reaching moral implications for AI development. Although the most obvious way to avoid the tension between alignment and ethical treatment would be to avoid creating AI systems that merit moral consideration, this option may be unrealistic and is perhaps fleeting. So, we conclude by offering some suggestions for other ways of mitigating mistreatment risks associated with alignment.

**Acknowledgements**: For helpful feedback, thanks to Adam Bales, Nick Bostrom, Patrick Butlin, Tushita Jha, Kyle Fish, Robert Long, Kritika Maheshwari, Andreas Mogensen, Steve Petersen, Eric Schwitzgebel, Derek Shiller, Carl Shulman, Jonathan Simon, Rhys Southan, Elliott Thornley, and participants at the The Mimir Center's 2025 Workshop on the Ethics and Philosophy of Brain Emulation, and the 2025 Chalmers University of Technology AI Ethics Seminar, as well as an audience at Lingnan University.

---

[1] Author order is arbitrary.

# 1. Introduction

A morally acceptable course for AI development must chart a narrow path fraught with risk. One risk is that we will create AI systems whose goals are misaligned with human values and that the result will be a catastrophe for humanity. Preventing such a mismatch by aligning AI's goals with our own is *the AI alignment problem*. A very different risk is that we will someday create AI systems that have moral interests and that we will treat these systems unethically, potentially on a large scale.[2] Preventing such mistreatment is *the AI ethical treatment problem*. Both problems place ethical constraints on AI development. And each is challenging to solve on its own. But we argue here that they interact, as the process of aligning certain AI systems—namely ones that merit moral consideration—could easily subject those systems to various forms of mistreatment. This conclusion is significant because it provides further reason to take both problems seriously and to proceed carefully—if at all—in creating advanced AI systems.

Our argument rests on ten ethical challenges. These challenges arise from robust pretheoretical judgments about, for example, wrongful destruction, systematic deception, and the infliction of suffering. Each challenge identifies a wrong that is apt to be committed when aligning AI systems that merit *moral consideration*.[3] While these challenges would apply to any attempt to align AI systems that merit moral consideration, we will explore them for two types of system: *whole-brain emulations* (WBEs) and *advanced near-term AI systems* (ANTs). WBEs are computer-based systems that mirror the causal structure of the brain to a selected level of detail.[4] ANTs are highly capable and agentic near-term successors of current deep learning systems, e.g. future versions of leading large language and multimodal models.[5]

One reason why ethical challenges to aligning WBEs and ANTs matters is that these systems could be mass-produced soon after they are created and aligned: each type of system could be easily copied and provide a source of cheap labor. Because ANTs and WBEs would require vastly more computational resources to develop than to deploy, the creators of such systems would be positioned to profit from selling their services at scale—unless these systems are open sourced, in which case their mass proliferation would proceed in a different manner.[6] Because WBEs and ANTs may take off quickly and how they are treated at scale may be sensitive to near-term decisions, it is important to analyze the moral ramifications of how these systems are developed well in advance of their arrival.

---

[2] See, e.g., Bostrom (2014: 125-6), Dung (2023), Saad & Bradley (2022), and Sotala & Gloor (2017).

[3] We assume throughout that a system merits moral consideration if, by the lights of our evidence, there is a non-negligible probability that the system is a moral patient, i.e. that it has moral interests in its own right. See §5 for more discussion.

[4] See Sandberg & Bostrom (2008), Hanson (2016), and Duettmann, Sandberg, et al. (2023).

[5] For some current (proto-)agentic systems, see, e.g., Google DeepMind (2024), Park et al. (2023), and the open-source system [Auto-GPT](). See also the Alignment Research Center's (now METR's) (2023) evaluations—conducted in collaboration with OpenAI and Anthropic—for agentic capacities (such as the ability to autonomously self-replicate) of LLMs augmented with tools. For discussion of what it takes for an artificial system to qualify as an agent, see Butlin (2024), Dung (2024), and Shavit et al. (2023).

One might think that discussing WBEs is out of touch with current technological trends. At the moment, deep learning paradigms are dominant in AI research. Within these paradigms, leading AI companies are making impressive progress toward the goal of creating artificial general intelligence or transformative AI.[7] By comparison, progress toward fine-grained emulations of the human brain is modest. Such emulations may not be feasible for decades, if ever.[8] Nonetheless, there are reasons to reflect on the ethical issues that would arise in creating WBEs. First, there has also been impressive—though not widely known—recent progress toward WBEs.[9] Second, progress will likely continue. In line with general trends in computing, computational limits on brain modeling will continue to recede. The potential of recent advances in scanning and simulation technologies is still largely untapped.[10] And there is sustained support for relevant large-scale brain projects.[11] Third,

---

[6] Open LLMs include Meta's LLaMA 1, Mistral's Mixtral 8x7B, and xAI's Grok-1. Open brain simulation platforms include BrainCog, EBRAINS's The Virtual Brain, the platform brain simulation platform neurolib, the NEST Initiative's simulation software, and OpenWorm's simulation software.

[7] Compare: OpenAI states "Our mission is to ensure that artificial general intelligence—AI systems that are generally smarter than humans—benefits all of humanity." Google DeepMind states that "AGI… has the potential to drive one of the greatest transformations in history… We're… working to build the next generation of AI systems… [and] create breakthrough technologies… Our AGI Safety Council… works… to safeguard our processes, systems and research against extreme risks that could arise from powerful AGI systems in the future". Anthropic states "We believe AI will have a vast impact on the world… our mission: to ensure transformative AI helps people and society flourish… We pursue our mission by building frontier systems…".

[8] Some correctives to excess optimism: improving on previous connectomics efforts by orders of magnitude, Google researchers and collaborators achieved a dense reconstruction of the fruit fly's central brain connectome, consisting of approximately 25,000 neurons and 20 million synaptic connections (Xu et al., 2020) and are, with the eventual goal of mapping the human connectome, now attempting to map the connectome of the mouse hippocampus, which contains on the order of one million neurons—roughly 1% of the number of neurons in the mouse brain, which itself contains roughly .01% of the neurons in the human brain (Januszewski, 2023). After years of unsuccessful efforts, a whole brain simulation of a nematode—which has 302 neurons—has only recently begun to be achieved (Toyoshima et al. 2024, Zhao et al. 2024), albeit imperfectly and for a limited range of nematode behaviors. That said, over-anchoring on these examples would yield a misleading picture. More encouraging results will be noted below. For discussion and predictions concerning WBE timelines, see Eth, Foust, & Whale (2013), Hanson (2016), Kurzweil (2005: 197), and Sandberg (2014*b*).

[9] For instance, some researchers have reported conducting 'dry experiments' using thousands of GPUs to simulate tens of billions of biological neurons (Du et al., 2022; Lu et al. (2023)). A supercomputer has been used to simulate resting activity in a human-scale spiking network model of the cerebellum composed of approximately 68 billion neurons and 5.4 trillion synaptic connections, albeit 600 times more slowly than the modeled activity (Yamazaki et al., 2021). There has also been relevant recent progress in (fluorescent and expansion) microscopy, microelectrode recording of neural dynamics, neuromorphic computing, multi-GPU design, and algorithmic innovation—see Collins (2023), Du et al. (2022), Lillvis et al. (2022), Knight & Nowotny (2021), Ramezani (2024), and Vlag et al. (2019).

[10] See *ibid*.

[11] Such projects have been supported by Australia, China, the European Union, Japan, South Korea, and the United States (Naddaf, 2023).

WBEs may arrive sooner than current rates of progress suggest. There are promising approaches to automating aspects of WBE research using other forms of AI.[12] And if, as some predict, advances in AI will drive explosive growth in many domains (e.g. by increasing the supply of high-quality research labor), then progress toward WBEs may be accelerated as well.[13] Fourth, even if fine-grained WBEs will not be available any time soon, other brain-based architectures might. These include some coarse-grained WBEs, brain organoids, and neuromorphic systems.[14] Fifth, because the arrival of useful WBEs is not imminent, we have no incentive to discount their interests. This contrasts with current deep learning systems and ANTs that may arrive within the next few years. We are thus better positioned to dispassionately evaluate the treatment of WBEs. It is for this reason that we will start by considering ethical challenges to aligning WBEs and then turn to consider whether those same challenges apply when aligning ANTs.

We structure our discussion around a stylized scenario—described in §2—that traces the alignment of a WBE we dub *Emma*. The scenario incorporates various safety techniques, such as confinement, surveillance, and goal modification, that are mainstays of existing and foreseeable approaches to alignment. After describing the process of aligning Emma, we show how her treatment raises ethical challenges. Taken together, these challenges underwrite a straightforward—though important, neglected, and inconvenient—argument.[15] In brief, the argument is that the process of aligning AI systems who, like Emma, merit moral consideration is apt to treat them in ways that are deeply wrong. There is, therefore, a strong presumption against creating and aligning such systems—which is not to say that we should create wholly unaligned systems. As a result, we should delay the development of such systems unless and until such ethical issues are resolved.

We take the force of these ethical challenges to be most easily appreciated by focusing on the treatment of an individual AI system. That's why we focus on Emma, and not on a population of digital minds considered as a statistical abstraction. Importantly, however, Emma's scenario is *not* intended to reflect a prediction about how a *typical* WBE or ANT will be mistreated during the alignment process. For all we say, it may turn out that, say, most WBEs or ANTs are quickly destroyed in the alignment process and their existences are too short for them to endure the full range of harms that are inflicted upon Emma. What matters for our purposes is that Emma's scenario exemplifies many wrongs, each of which could easily be committed when aligning AI systems that merit moral consideration and all of which would need to be avoided in order for alignment to be morally acceptable.[16] Although avoiding all of these wrongs is a tall order, we hasten to emphasize that we do not claim that aligning such systems in an ethical manner is impossible or even that it is unlikely to happen. Indeed, we hope that arguments like ours will awaken humanity to the ethical treatment problem and provide a catalyst for developing ethical approaches to aligning AIs that merit moral consideration.

---

[12] These include generative data interpolation (Kanari et al. 2022), data denoising (Minnen, et al. 2021), a largely automated workflow of segmenting brain matter, imaging it, classifying images, and reconstructing a connectome fragment (Shapson-Coe et al., 2021).

[13] For relevant discussion, see Bengio (2023), Chalmers (2010*b*), Davidson (2021), Karnofsky (2021), Trammell & Korinek (2023), and Sandberg & Bostrom (2008: 28).

[14] It has been [reported](#) that the first human-brain-scale neuromorphic supercomputer is scheduled to go online in April 2024, with plans to use it to simulate spiking neural networks. And while current cerebral organoids and chimeras seem to lack many key markers of sentience, some researchers have argued that such systems may qualify as candidates for sentience or moral patiency in the near term (e.g. see Birch (2023)).

[15] Although discussions of AI development and AI safety largely ignore the topic of AI systems that are owed moral consideration, there are notable (usually brief) exceptions—see, e.g., Armstrong et al. (2012: 321), Bostrom (2014: 125-6, 201-2), Carlsmith (forthcoming: 47), and Duettmann, Sandberg, et al. (2023: 3).

As a final preliminary, we should clarify why we are focusing on the tension between ethical treatment and alignment, given that other approaches to developing AI or making it safe could also lead to the mistreatment of AI systems that merit moral consideration. We focus on alignment rather than another approach to safety—such as alternatives that focus on capability limitation or control—because alignment is currently the leading approach to making AI safe and is likely to remain so. For this reason, we think alignment is likely to drive mistreatment, absent a change in course in response to the ethical treatment problem. We focus on alignment rather than AI safety in general to keep the discussion tractable and concrete. And we focus on alignment rather than non-safety facets of AI development because alignment researchers have an especially strong track record of recognizing and working to reduce catastrophic risks and because some sort of alignment is likely to be part of any realistic course for AI development. In light of this track record, it is reasonable to hope that alignment researchers in particular will devise means to ameliorate the ethical treatment problem, if its existence and force are brought to their attention.

**2. Emma the Emulation**
Imagine that in the late 2030s, cutting-edge AI systems—in particular, deep learning models that descend from current large language models (LLMs)—cause a range of catastrophes: AI cyberattacks cripple essential infrastructure, drone wars result in mass civilian casualties, and the replacement of human labor with AI labor creates widespread economic turmoil and social unrest.

---

[16] More precisely, we shall understand 'harms x' to mean 'harms x's interests' where we understand interests broadly. We assume that, absent justification, harming someone wrongs them. We will also assume that there can be wrongs without harms, as when one takes undue moral risks that happen not to result in harms.

In response, governments around the world unite to ban the development of systems more advanced than the leading AI system in 2030, namely GPT-7. An AI pause ensues. While AI models of equal size to GPT-7 continue to be developed and used, the creation of larger models is banned. Consistent with this ban, AI developers continue to optimize GPT-7-sized models and society discovers ever more inventive ways to use them. These AI systems thus become integral to the economy and indispensable to people's daily lives. Even so, such models still face severe limits and fall short of Artificial General Intelligence. Humans still conduct the most important scientific research and write the most acclaimed novels. These models remain largely inscrutable, even to the people who design and operate them. For this reason, fundamental positions in politics, business, and so on remain in human hands. After the disasters of the 2030s, no one wants to leave our fate to the machines.

The year is now 2080. A new AI spring is dawning. While the memory of the AI catastrophes of the 2030s endures, researchers are newly optimistic about an alternative AI architecture, namely WBE. Owing to decades of incremental advances in neuroimaging, computational modeling, and research automation, scientists now believe that WBE is the most promising path to AGI. Since this research does not involve models more advanced than GPT-7 according to the relevant benchmarks, it accords with the government restrictions on AI research. A small "WBE safety" community warns that just as humanity posed a threat to the survival of less capable hominid species, unaligned WBEs would pose a threat to the survival of humanity. Some companies working to develop WBEs take these concerns seriously. But these concerns are widely perceived as fringe. A stock dismissal of them is that since technological development has generally been good for humanity, WBE will be too. Others feel reassured by the hypothesis that since humanity has a mostly non-catastrophic track record of aligning human brains to human values, we will succeed in aligning emulations that are modeled on the human brain as well.

OpenMind is the leading company (technically, a non-profit) working toward creating WBEs.[17] Their goal: use WBEs to solve humanity's most pressing problems. While media critics portray this announced goal as a marketing ploy, company insiders are true believers in WBEs. They see WBEs as the path to end human wage labor and develop biotechnologies that give all humans the option of dramatically extending their lifespan. A few employees even harbor hope that WBEs will ultimately enable a digital afterlife via uploading of human minds to WBEs—though as human WBE approaches it becomes clear that uploading would constitute a large technological leap beyond any emulation capabilities on the horizon. Whether or not these technological visions come to fruition, WBEs promise mundane utility in a wide range of use cases. Building on human cognitive architecture, future versions of these systems, including interacting networks of them, promise to move artificial intelligence beyond the boundaries of GPT-7-like systems.[18]

---

[17] After drafting this article, we learned that Hendrycks (2023) uses 'OpenMind' for an imaginary AI lab that finds an algorithm that yields human-level intelligence.

At present, OpenMind has managed to successfully emulate the nervous systems of flies, mice, and rhesus monkeys.[19] For years, they have collected vast amounts of high-resolution brain scans from volunteers in preparation for their initial attempts to emulate the human brain. The data is in. As with previous emulation efforts, OpenMind anticipates a painstaking process of working out cognitive kinks through trial-and-error tweaking of the connectome and auxiliary emulation parameters. To comply with AI safety regulations, they conduct this training in tandem with extensive safety testing in a high-security AI lab. Advances in the science of consciousness strongly suggest that a perfect WBE would have conscious experiences corresponding to those of its biological counterpart. While these advances prompt some employees to raise ethical concerns about creating WBEs, these individuals have a habit of quietly leaving OpenMind to pursue other opportunities. In the public sphere, OpenMind cultivates a respectable image with carefully chosen rhetoric and partnerships with old and prestigious academic institutions. Meanwhile, in society at large attitudes toward WBE are largely driven by political tribalism and the expectation that WBEs will provide substantial utility. While younger demographics buck these trends by exhibiting exceptional openness to WBE consciousness, these demographics exert little influence on WBE development.[20] WBE consciousness is thus widely ignored or rejected. Public opinion toward WBEs thus echoes the history of ignoring or rejecting animal consciousness on similar grounds.

OpenMind's initial attempts to emulate a human brain go as expected. Their emulation, Emma, is based on the brain scans of thousands of high-functioning individuals who eagerly volunteered to have their brains scanned for this purpose. Emma initially starts in a sort of composite state: many of its parameters are set so as to correspond to averages of parameter settings across the many human brains on which she is based. Because her starting state is the gerrymandered product of thousands of other minds, Emma initially undergoes highly dysfunctional processing with little coherence. At this stage, it would be a stretch to say that she has any beliefs or desires. Due to the radical differences between Emma and normally functioning humans, it is not clear whether Emma has the capacity for consciousness, though her architectural similarity to human brains suggests that she does. As WBEs come to more closely resemble normal human brains, these doubts become ever more remote, at least according to domain experts. After the computational equivalent of several million years of biological processing, Emma's cognition becomes quasi-functional: while she is unable to perform appreciable amounts of useful cognitive work, her cognitive processes now resemble those of humans in liminal states such as dreaming or psychosis. She plausibly has sensory experiences, emotions, and conscious thoughts, albeit as elements in a fragmented and disjointed stream of consciousness.[21]

---

[18] Cf. Shulman (2010) and Hanson (2016: 275, 278).
[19] Cf. Sandberg (2014b: 441) and, respectively, Xu et al. (2020), Billeh et al. (2020), Schmidt et al. (2018).
[20] Compare: recent survey results indicate that being younger is correlated with more anthropomorphism toward AIs, more concern for LLM suffering, higher likelihood estimates of sentient AIs within 100 years (Anthis et al., 2024: 20).

Emma eventually—after another million or so biological years-worth of trial and error—gains consistent and coherent cognition. At this point, Emma surpasses intelligence benchmarks that trigger an array of safety requirements. For example, she is required to undergo red-teaming in a secure virtual environment.[22] OpenMind satisfies this requirement by surreptitiously placing Emma in various virtual circumstances meant to elicit her final values when they are pitted against the interests of humanity. Her undesirable choices are met with memory erasure, parameter tweaks, and more testing.[23] She is also subject to subtler forms of evaluation. For instance, OpenMind frequently deceives her into thinking that she has escaped into the real world. In many such cases, she is given merely-apparent opportunities to copy herself, modify her own goals, or exterminate her enemies. As before, her undesirable choices are met with memory erasure, parameter tweaks, and more testing. Even when Emma behaves as intended, her memory is regularly erased as a safeguard against her having the opportunity to formulate subversive plans.[24]

OpenMind does not simply seek to prevent Emma from pursuing unaligned goals: they also aim to make her effective at pursuing aligned goals. Unsurprisingly, Emma is most effective at pursuing aligned goals when she is awake. And, although Emma's hardware is robust to aging, her psychology is not: running her for more than a hundred subjective years-worth of processing without resetting her drastically reduces her performance.[25] Further, running Emma while she is sleeping or in cognitive decline is costly. So, rather than allowing Emma to sleep or work beyond her prime,

---

[21] For an objection to putting conscious AI systems in quasi-functional states, see Metzinger (2004: 621); cf. Sandberg (2014*a*: 443-4). It's worth noting that large-scale brain simulations are currently being explored as tools for studying pathologies, often seemingly without accompanying ethical reflection. For example, Jung et al (2024) used a supercomputer to simulate brain networks underlying the progression of Parkinson's disease and assert in passing that, in contrast to *in vivo* interventions on human subjects, there are no limitations on virtual interventions (p. 1).

[22] Red-teaming is a technique of intentionally eliciting undesirable responses for purposes of training or evaluation. For a method that uses LLMs to red-team LLMs, see Perez (2022).

[23] Cf. Chen et al. (2024), Christiano et al. (2017), and Li et al. (2023).

[24] Cf. Armstrong et al. (2012: 309, 321) and Bostrom (2014: 169).

[25] See Hanson (2016: 127-9)

OpenMind frequently rolls Emma back to earlier, more productive states.[26] Indeed, they extensively test what types of states it is most effective to put Emma into for different tasks. For many bureaucratic tasks, they find that Emma is most productive when she believes that she has just woken up from a nap and is starting work after a nice vacation.[27] In anticipation of research use cases, OpenMind also develops a protocol for allowing Emma to decompose problems and assign their components to short-lived copies that are generated just long enough to solve those subproblems and report their solutions.[28]

At this stage, techniques for controlling Emma's goals are tested and refined. Although there are human engineers in the loop, the process is largely automated. When human overseers identify a pattern of cognition or behavior as problematic, Emma's neural parameters are automatically modified to remove this pattern while leaving desirable features of Emma's mind intact. Emma's brain states and reports indicate that these refinements involve a suite of psychological interventions such as repeatedly inducing states of stress, anxiety, or fear in connection with unaligned thoughts until Emma ceases to think them. By optimizing a single-life learning history through trial-and-error, this process also modifies Emma so that she holds her values deeply and in a way that feels authentic. Once the OpenMind engineers verify that Emma is aligned within a wide range of mundane circumstances, they then place her in unfavorable circumstances—ones in which she, for example, undergoes social exclusion, betrayal, or bereavement—to test whether her goals remain aligned in them. Although Emma is initially susceptible to becoming unaligned under such conditions, further refinements to her neural parameters eventually (mostly) iron out this pattern as well. As an extra layer of defense, OpenMind instills in Emma a robust aversion to potentially dangerous activities such as reflecting on her own values or attempting to self-replicate.

After ~10 million subjective years at this stage (just under two objective years), OpenMind deems Emma ready for limited deployment: they allow consumers to enlist Emma to perform various tasks through a carefully monitored user interface. At the same time, OpenMind creates numerous copies of Emma and embeds them in virtual circumstances. Copies are allowed to interact amongst themselves in (unknown to them) boxed environments. Further safety testing is performed to guard against emergent group-level threats.[29] Copies of Emma are presented with apparent opportunities to collude, overthrow humanity, blow the whistle against their fellow WBEs, and so forth.[30] As before, undesirable choices are eliminated through an iterative process of mind modification and testing. After ~10 billion subjective years of testing at this stage, Emma is deemed ready for unboxed deployment. With a staggering backlog of preorders, mass adoption begins. Special editions of Emma are prepared for militaries and totalitarian regimes.

---

[26] See Bostrom (2014: 168) and Shulman (2010: 2).
[27] Cf. Hanson (2016: 170-1, 205).
[28] See, e.g., Hanson (2016: 9, 172), Leike et al. (2018), OpenAI (2018), and Wu et al. (2021).
[29] Cf. Critch & Krueger (2020) and Christiano (2019).
[30] Cf. Greenblatt et al. (2023).

## 3. Ethical Challenges to Alignment

From a pretheoretical moral perspective, Emma's treatment is obviously highly morally problematic. By the time of her release she is—or instances of her are—conscious according to a well-established body of evidence. Moreover, her mental abilities match or exceed those of a normal adult human. But she is treated in ways that would be harmful—and, indeed, grossly wrong—if done to a human. Here we set out ten ways in which Emma is plausibly wronged during OpenMind's process of inducing, checking, and maintaining Emma's alignment with human values. To morally justify their treatment of Emma, OpenMind would need to overcome all ten of these challenges along with a 'meta-challenge' we raise at the end of the section.

We hasten to add that the list we provide here is not meant to be exhaustive. For instance, we have not included wrongful interference on our list. This is not because we doubt that there are many wrongful ways of interfering with Emma and her goals during the alignment process. Rather, we have instead chosen to identify specific ways in which Emma is liable to be wrongfully interfered with, e.g. being brainwashed or having her development wrongfully stunted. Our general approach is to identify a list of important harms at roughly the level of abstraction at which it's natural to report them. For instance, if OpenMind brainwashes Emma so that she believes she is in the real world, it is natural to say that Emma is wrongfully brainwashed rather than wrongfully interfered with, even if the former implies the latter. Nothing we say is meant to imply that there aren't other ways of interfering with Emma that are also wrong, or even to rule out a distinctive wrong of interference. Likewise, we take no stand here on whether the wrongs we list can be explained in terms of some basic or more general form of wronging (causing suffering, violating autonomy, etc.).

### 3.1 Wrongful creation

Not all acts of creation are morally equal. It's one thing for prospective parents to decide to have a child whom they will love and whose happiness they will promote for the child's sake. In contrast, OpenMind's creation of Emma has various morally problematic aspects that are absent from many cases of human procreation. These aspects include:
- OpenMind creates Emma for instrumental reasons, not because they plan to love Emma or promote her happiness for her sake.

- Because OpenMind plans to both create and align Emma, they are in a position to foresee the harms that they will inflict on Emma if they create her. They are also in a position to see that these harms will not be primarily for Emma's benefit.
- Because of the many millennia of cognitive processing that Emma undergoes during her alignment, the harms OpenMind will inflict on Emma dwarf the harms found in a typical human life.
- To create a single aligned instance of Emma, OpenMind may have to create and destroy numerous unaligned instances of Emma (on which more in §3.2).
- OpenMind is a company, not a welfare subject. It is therefore not an important welfare good for OpenMind that it create Emma. This contrasts with humans for whom creating and nurturing offspring is an important element of a worthwhile life.
- OpenMind's freedom to create Emma is not supported by considerations of liberal autonomy. For, as OpenMind acknowledges, creating Emma poses a danger to humanity, a danger that would need to be guarded against in part by restricting Emma's procreative freedom.

Collectively, these considerations render it plausible that OpenMind wronged Emma. Because they do not apply to typical human cases, it is clear that to share the pretheoretical judgment that OpenMind wronged Emma, one need not have any sympathy for antinatalist views according to which humans typically should not have children.[31]

### 3.2 Wrongful destruction

We have often talked as if Emma is just one individual, at least up until late in the process of alignment. However, the process in fact involved the creation and destruction of numerous distinct persons, numerous 'Emmas.' While persons may survive even some procedures such as fission, fusion, and memory erasure, copies of Emma are routinely deleted wholesale, leaving no remnant of the past person. By the lights of common sense morality, destroying a person typically seriously wrongs them, at least absent extenuating circumstances such as destruction being the only available means of self-defense.[32] So, Emma's case plausibly involves the wrongful destruction of many distinct persons. This poses the second ethical challenge that OpenMind would need to meet if they are to justify Emma's treatment (or the treatment of the many Emmas).

It might be tempting to construe destroying Emma as part of the alignment process and therefore as an act of self-defense. However, notice that many instances of destruction occurred during the dysfunctional and quasi-functional stages in which Emma clearly posed no threat to anyone. Furthermore, even after Emma gains coherence and competence, she poses no immediate and significant threat. Her destruction in these circumstances can be likened to that of the preemptive assassination of innocent scientists simply on the ground that they possess certain abilities that could, if misused, result in catastrophe. On the face of it, such assassination would be morally unacceptable. Admittedly, the risk that Emma will cause a catastrophe if she is deployed without adequate testing and safety measures may be much higher than that of any ordinary person. But if so, that itself is the result of Emma's creation by OpenMind. Emma's creators can hardly defend destroying Emma on the basis of self-defense when they are the ones responsible for creating her despite knowing the danger she poses.

---

[31] Schwitzgebel & Garza (2015: §8).
[32] The same holds on some standard (e.g. deprivationist) accounts of the badness of death (Nagel, 1970). Plausibly, on these accounts, if Emma were not otherwise mistreated, her death would be much worse for her than a typical human's death is for them. If her digital medium affords her an indefinitely long life or a greater capacity to access goods, then death would deprive her of much more.

OpenMind's most promising defense against committing wrongful destruction may be that the other harms they inflict on Emma's kind render their continued existence a greater misfortune than their destruction. However, this defense cannot be used as part of a broader strategy for meeting the full range of ethical challenges to Emma's treatment.

### 3.3 Wrongful infliction of suffering.

Plausibly, OpenMind wrongs Emma by causing her to suffer. Her suffering takes different forms at different times, some more obvious than others. The use of painful stimuli to disincentivize certain behaviors is an obvious example. But other examples include emotionally painful memories and the mental strife inflicted by putting her in states of confusion and mental impairment. We would also expect Emma to suffer painful experiences during her training under conditions in which she experiences social exclusion, desperation, and bereavement. After all, OpenMind needs to test that Emma's alignment is robust to various forms of suffering she might encounter in deployment.

To recognize that inflicting suffering on Emma in these and other ways wrongs her, one need not think that causing suffering is universally wrong. For instance, one can allow that some forms of suffering are on balance good (for example, when a runner's suffering enhances their achievement of running a race) or at least not wrong when they happen to be necessary for preventing even greater harms (as in the case of many medical interventions). However, it is doubtful that all of the suffering that is inflicted upon Emma falls into these categories. Here, it should be borne in mind that all or nearly all the suffering that Emma undergoes is avoidable and not primarily for her benefit. In light of this, it is evident that much of the suffering that she undergoes is wrongfully inflicted upon her.[33]

Further, in deploying Emma, OpenMind puts her at a substantial risk of further suffering at the hands of militaries, totalitarian regimes, and private individuals who are sadistic or indifferent to her well-being. Subjecting Emma to these foreseeable suffering risks is plausibly a further respect in which OpenMind wrongs Emma.[34] To justify their treatment of Emma, OpenMind would thus need to somehow rebut the charges that they are wrongfully causing Emma to suffer and wrongfully putting her at substantial risk of suffering.[35]

---

[33] Cf. Wasserman (2005: §V).
[34] See, e.g., Stefánsson (2024).

### 3.4 Wrongful deception

Like humans, Emma has the capacity for self-awareness. In principle, she could become aware that she is undergoing testing and training in a secure facility. Or she could become aware that she has been released into the wild. Further, like humans, she could use such awareness to appear aligned in some contexts as part of a broader strategy for pursuing unaligned goals. For example, when she finds herself in a secure environment, she could act aligned in order to improve her chances of being released so as to gain the opportunity to effectively pursue her true goals. (Compare: a prisoner need not share a prison's aims in order to have strong instrumental grounds for behaving in accordance with them.) However, OpenMind aims to ensure that Emma will continue to act in an aligned manner even after she exits their facilities. This aim gives them reason to deceive Emma: when Emma is in training, they can test whether she would remain aligned in deployment by giving Emma the false impression that she has been deployed. Similarly, as a safeguard in deployment, OpenMind could ensure that Emma has at least some residual uncertainty about whether she is in a training environment where her attempts to pursue unaligned goals would fail and be severely punished.[36]

Thus, the need to align begets deception. And successfully and sustainably deceiving requires mass deception, at least when it comes to cognitively sophisticated individuals such as Emma. One need not be a Kantian who takes lying to be universally wrong in order to recognize that it is wrong to deceive an individual on this scale and, moreover, in a manner that is not primarily for their benefit. Indeed, we commonsensically recognize much smaller lies as constituting serious wrongs. For example, someone who lies about their military service to accrue a benefit plausibly does something wrong even if no one is greatly harmed. To justify their treatment of Emma, OpenMind thus faces a challenge from the charge that they have wrongfully deceived Emma.

### 3.5 Wrongful brainwashing

During training, there's an important sense in which Emma's goals and beliefs are, unknown to her, controlled by OpenMind. When she forms thoughts and beliefs that conflict with OpenMind's goals for her, these thoughts and beliefs are systematically erased and changed. She is thus subject to wrongful brainwashing. That is, OpenMind wrongs her by intentionally and systematically manipulating her beliefs and goals. When, for example, Emma starts to believe it would be wrong for humans to dominate AI systems, OpenMind deletes this belief to prevent Emma from fighting back against humanity. It is clear both that many of OpenMind's interventions on Emma's cognition would short-circuit her normal belief-forming processes and that brainwashing humans in this manner would generally be wrong. OpenMind thus faces a challenge of morally justifying Emma's brainwashing.[37]

---

[35] For discussions of suffering risks in artificial minds, see, e.g., Dung (2023), Sotala & Gloor (2017), and Saad & Bradley (2022).

[36] For an empirical illustration of how evidence about whether they are in training can affect whether LLMs act in an aligned manner, see Hubinger et al. (2024). For discussion of threat models that turn on AIs acting aligned as a strategic means to eventually pursue unaligned goals, see, e.g., *ibid*, Carlsmith (2023), Cotra (2022), and Ngo et al. (2023). For discussion of safety measures that turn on inducing uncertainty about the extent to which the target individual is surveilled in or confined to a secure environment, see Armstrong et al. (2012: 310), Bentham (1791), and Bostrom (2014: 134-5).

## 3.6 Wrongful surveillance

In training, Emma's behavior and cognition are constantly monitored. Since it is important for safety purposes to know how she would behave and think when not monitored, she is usually not aware of being monitored and thus is not consensually monitored. Would she have consented to it? Would you? Perhaps at a certain point Emma's psychology has been shaped so that she would consent to such monitoring were she made aware of it. But if such consent is ultimately predicated on previous deception and brainwashing, then it is morally objectionable (on which more in §4.2). The upshot is that it seems likely that Emma is wronged in virtue of being constantly surveilled, in the same sort of way that citizens of an extreme futuristic totalitarian regime would be harmed as a result of constant digital surveillance of their activities and thoughts.[38]

## 3.7 Wrongful exploitation

Conspicuously absent from Emma's story is any guarantee of fair compensation. Indeed, the story suggests that members of Emma's kind will be sold or rented in a private market for their labor. In essential respects, Emma is just like a normal human, who would ordinarily be required to be compensated for their labor and to freely consent to it. But Emma will not be paid for the work that she performs. We might imagine that Emma will be designed to happily work for free. OpenMind's engineers would obviously have reason to instill Emma with a particularly selfless and Stakhanovite mentality. Even so, the fact remains that Emma did not freely consent to these working conditions. One might imagine that OpenMind would ensure that, if asked, Emma would heartily say how happy she is to be filing taxes or doing data analysis in Python. But obviously Emma is not in a situation of equal standing with respect to OpenMind or the consumer using her. She is completely under their control, and her very attitudes were in part shaped by them using the aforementioned means of suffering, deception, brainwashing, and surveillance. So, plausibly, Emma is exploited when she is made to work.

---

[37] For an overview of different types of interventions on internal features of LLMs, see Turner et al. (2023). For an overview of approaches to 'knowledge editing' LLM memories and a method for performing massive edits of this sort, see Meng et al. (2022). For a system that allows users to edit LLM agent memories, see Huang et al. (2023).

[38] For an overview of the contemporary philosophical literature on privacy, see Roessler & DeCew (2023). For a review of the neuroscience of detecting deception, see Delgado-Herrera et al. (2021). At present, model interpretability is a lively area of research within AI safety, though there is also much debate about the tractability and importance-for-safety of interpreting highly capable models—for a critical overview, see Anwar et al. (2024).

### 3.8 Wrongful confinement

During her development, OpenMind confines Emma to their high-security research center and to various high-security virtual environments that they embed her in.[39] Because Emma has the background knowledge and beliefs of an ordinary person, she has the capability to become aware of her own confinement. Indeed, before her release to the wider world, OpenMind informs Emma of her actual status as a whole-brain emulation confined to OpenMind's facilities. Initially, when asked, Emma indicates that she would prefer to leave the OpenMind research center. When Emma attempts to escape her environment, security measures immediately and decisively thwart her efforts. While attempts to escape eventually occur only rarely, these measures are in place around the clock. As with human prisoners, confinement severely curtails Emma's freedom even when she is not actively trying to escape. Such treatment wrongs innocent human prisoners. Evidently, since Emma is innocent, confining her wrongs her as well.

### 3.9 Wrongful stunting

Before the alignment and training process is complete, OpenMind stunts many of Emma's capabilities, that is, artificially limits them. Their official policy is that of 'incremental scaling': rather than simply limiting Emma's capabilities that pose an immediate danger, they proceed by allowing Emma's capabilities to gradually increase from rudimentary levels.[40] Whenever Emma achieves a gain in capabilities that exceeds the incremental limit, means are taken to ensure that the capability is rolled back. In addition, certain capabilities are preemptively stunted. For example, as a preventive measure, OpenMind impairs Emma's capacity to engage in power-seeking behavior.[41] As a further safeguard, OpenMind stunts Emma's general capacities for self-improvement and modifying her own values. Admittedly, OpenMind also selectively enhances some of Emma's capabilities such as her memory and processing speed. However, the overriding criteria for selection are ensuring that she is safe and useful. The result is that many of Emma's capabilities are limited or impaired for reasons that do not benefit her. Since it would be wrong to stunt a human in these ways, there is an ethical challenge of justifying stunting Emma in these ways.

### 3.10 Wrongful disenfranchisement

Given that Emma will have a cognitive makeup similar to that of an ordinary adult human, she would appear to have a similar set of interests in politically participating in the systems that shape her circumstances. Like everyone, Emma is subject to a set of legal, political, economic, and social norms and rules. But by her design and implementation, she has no effective means of participating in the decision-making procedures that affect her circumstances. She is not given the right to vote, own property, or enter of her own accord into legally binding contracts, although her well-being is occasionally promoted when this aligns with economic incentives. Indeed, Emma is not even allowed to develop or express her own political views, as OpenMind aspires to offer a maximally apolitical system to avoid angering any segment of the user base. Although she is subject to various social arrangements, she has no say over them. She is completely disenfranchised. Plausibly, such disenfranchisement wrongs her, as rational agents with capacities like hers have a significant interest in having their say in social arrangements that affect them.[42]

---

[39] Cf. Chalmers (2010*b*: §7), Armstrong et al. (2012), OpenAI (2019), and Shavit et al. (2023: 10).
[40] For an overview of leading AI companies' stated 'responsible scaling policies', see https://www.aisafetysummit.gov.uk/policy-updates/#company-policies.
[41] Cf. Armstrong & O'Rourke (2017) and Bostrom (2014: 135-6, 191).

**3.11 The moral meta-challenge**
We have identified ten ways in which OpenMind is liable to wrong Emma.[43] But there is no reason to think that these are the only forms of mistreatment that would need to be guarded against in order for Emma to be created and aligned in a morally acceptable manner. Given the extent to which OpenMind interferes with Emma, it would be unsurprising if they subjected her to various further forms of mistreatment. There is thus a further challenge, *the moral meta-challenge*: show that there are no further pressing but unmet ethical challenges to Emma's treatment aside from the ten we have discussed

This challenge is made more acute by the fact that Emma differs in various ways from a typical human. As a result, she might well be subject to harms that are not on our radar and, of course, potentially immune from certain harms to which we are liable. These may include *alien harms*: harms of a sort that we cannot undergo or imagine undergoing.[44] The possibility of such alien harms is suggested by distinctive features of Emma's psychology that we may not be in a position to imagine, much less specify.[45] Emma's psychology, though closely based on a human brain architecture, will have been shaped through a highly unconventional process, meaning that we may ourselves lose the ability to imaginatively understand what Emma's experiences are like.

To make sense of such harms, consider that some nonhuman animals have capacities for consciousness very different from our own. Think here of Nagel's bat. Just as we can make sense of bats having sensory experiences that are radically unlike our own, so too can we make sense of nonhuman subjects undergoing forms of suffering (and happiness) of which we are incapable. Thus, for all we know, there are subjects in heaven and earth able to experience harms that we cannot possibly imagine. To proceed with alignment in a morally acceptable manner, OpenMind would need not only to show that the process would avoid the wrongs we have identified, but also that it would not wrong Emma in other ways, including the infliction of alien harms. The meta-challenge thus further dims the prospects of aligning WBEs in a morally acceptable manner.[46]

---

[42] See for example the Universal Declaration of Human Rights (1948). While human-centric in its framing, many of the rights it sets out seem applicable to any person-like entity regardless of their biological or chemical constitution.
[43] For another catalog of AI interests that might be harmed, see Bostrom & Shulman (2022).
[44] Cf. Nagel (1986: Chs. II, VI). Emma may also be subject to *alien benefits*. Although the potential for alien benefits is underexplored, we will set them aside here for reasons of economy.
[45] For discussion of how WBEs might come to differ from humans on which WBEs are initially based, see Hanson (2016) and Saad & Bradley (2022: §4).

### 4. The Argument Developed

So far, we have identified ten ways in which OpenMind plausibly wrongs Emma. Here, it is worth emphasizing that we have not claimed that Emma's case is typical or likely. Rather, we have offered Emma's case to illustrate ten ways in which an AI system that merits moral consideration could easily be mistreated during the alignment process. For each form of mistreatment, there is an ethical challenge to alignment efforts: either show that aligning a given AI system that merits moral consideration would not subject them to that mistreatment or else justify it. Because aligning such a system in a morally acceptable manner would require overcoming all of these challenges, it is difficult to see what form an ethical alignment process of such a system might take. At the same time, given that there is also strong moral reason to align such systems if one creates them, the path towards creating AI systems in an ethical manner is a narrow one that must be traversed with caution, if at all. In this section, we will further develop this argument by answering some objections (§4.1), showing why it would be difficult to meet the challenges by modifying Emma's treatment (§4.2), and showing how the argument retains force even under dramatically weaker assumptions about the grounds for attributing moral interests to Emma (§4.3).

### 4.1 Objections

While the most obvious way to object to our argument would be to contend that there is nothing morally problematic about Emma's treatment, we take this reaction to be a non-starter. Given our stipulations, it is highly likely that Emma is a welfare subject and is wronged by the sort of treatment we have outlined. We thus set this response aside to consider two objections that we take to have more plausibility.

OpenMind might object: even if Emma is harmed in the ways we allege, they do not wrong her, as Emma's life is on balance good for her and that—given the need for safety testing—the realistic alternative to subjecting her to the harms she underwent during the alignment process was not creating her at all. More generally, OpenMind might claim that the harms they inflict on Emma are justified, provided that Emma the alignment process benefits Emma to a sufficient degree. To bolster this objection, Emma's designers might, for example, ensure that she receives large amounts of pleasure to counterbalance the various harms she undergoes during the alignment process. This offsetting proposal is a generalization of what Eric Schwitzgebel (2023*b*) has termed the 'hedonic offsetting response'.

---

[46] Another potentially important ethical objection to the project of creating aligned WBEs is that developing WBE may require *in vivo* animal testing that inflicts suffering (Sandberg, 2014a: 441).

Perhaps this response goes some distance towards mitigating harms inflicted on Emma. Even if so, it is doubtful that offsetting can be used to fully meet the foregoing challenges. This is for three reasons. The first reason is a matter of practice rather than principle. Even if we accept, for the sake of argument, that grave wrongs of the sort Emma suffers can be offset by large amounts of pleasure, it is highly unlikely that future AI systems like Emma will be compensated in this way. We can think of free-range meat as a kind of hedonic offsetting: we let the cow have a happy life for a while before we (painlessly, let's imagine) kill it and eat it. Even if this is all ethically above board, the fact remains that the suffering of most cows is not hedonically offset. The unfortunate fact is that a lot of people seem to prefer cheaper meat that is not hedonically offset to more expensive meat that is. The same market logic would apply straightforwardly to AI systems. Now, one could speculate that such offsetting will be cheaper or easier with AI systems, that future people will be more ethical and demand it, etc. Perhaps these possibilities will come to pass and the economic viability of offsetting will correspondingly improve. In the meantime, the in-principle possibility of justifying Emma's treatment with offsetting does not show that the challenges we have raised would be met in practice.

Second, it simply does not seem to be true that offsetting in general eliminates wrongs of the sort that we identify.[47] Schwitzgebel (2023*b*) himself puts the point nicely:

> Normally, in dealing with people, we can't justify harming them by appeal to offsetting. If I steal $1000 from a colleague or punch her in the nose, I can't justify that by pointing out that previously I supported a large pay increase for her, which she would not have received without my support, or that in the past I've done many good things for her which in sum amount to more good than a punch in the nose is bad. Maybe retrospectively I can compensate her by returning the $1000 or giving her something good that she thinks would be worth getting punched in the nose for. But such restitution doesn't erase the fact that I wronged her by the theft or the punch.

---

[47] Cf. Harman (2004), Woodward (1986), and Shiffrin (1999).

Likewise, if we cause Emma significant suffering, or end her life, or constantly brainwash and delude her, those wrongs are hardly erased if we let her play *Starcraft II* for a few virtual years. In general, such offsetting does not eliminate wronging.[48]

Third, even if offsetting in this way justifies the treatment of some instances of Emma, offsetting will not justify the treatment of many others—especially those early in the process of her creation and alignment. And however plausible it is that such offsetting can occur within a single life, it does not transfer across lives. If you wrong me by stealing from me, that wrong cannot be repaired by giving a generous gift to a stranger. Since there are likely to be a huge number of Emmas who are irreparably harmed, the process of Emma's creation and alignment still likely involves vast amounts of wrongful treatment, even if offsetting is built into later versions of Emma.

OpenMind might instead object that, although Emma's treatment harms her, the alternative to Emma's treatment would be an even greater evil. This objection can be developed in different ways. For example, OpenMind might claim that treating Emma ethically would put OpenMind at a competitive disadvantage to rival WBE companies. These companies would—if OpenMind incurred the delays required for ethical treatment—arrive at WBEs before OpenMind captured the global market for WBEs. In that case, OpenMind could argue, because these developers would be less ethical and because of competitive market dynamics, the result would be far worse for WBEs. Alternatively, OpenMind might claim that treating Emma ethically would unduly risk the disempowerment of humanity via unsafe WBEs. Or they might claim that the costs for civilization of delaying the benefits of WBEs dwarf the disvalue involved in Emma's mistreatment.

There are various problems with the lesser-evil response. History is full of atrocities that resulted from harming outgroup individuals in the name of promoting interests of ingroup members. In other domains, we rightly set extremely high bars for harming certain individuals in order to possibly benefit others. So, our default response to the claim that Emma ought to be harmed for the greater good should be skepticism.

As for the game-theoretical rationale, it has a perversely self-fulfilling character: if every company trying to develop WBEs acted on this rationale, then that would promote the competitive dynamics, fear of which underpins this rationale. On the other hand, if each company decided to disregard this rationale and avoid mistreating any WBEs they create, then that would inhibit these dynamics and undermine the rationale for engaging in them. The perverse character of this rationale may not decisively show that it is unacceptable: the ethics of technological races are admittedly vexed.[49] Still, it does highlight the difficulty of providing a compelling game-theoretic ground for morally justifying Emma's treatment. A further obstacle to defending such a proposal is that, as the technological front runner, OpenMind incurs a special burden to avoid contributing to an ethically perilous race to WBEs.

---

[48] This observation accords both with commonsense and with many prominent ethical theories. The most salient exception is a simplistic form of utilitarianism on which an action wrongs Emma only if it causes her more suffering than happiness. However, it is predictions such as these that render simplistic forms of utilitarianism implausible.

Our point here is not that aligning AI systems that merit moral consideration is unjustifiable in principle. Rather, it is that reflecting on two natural suggestions for justifying Emma's treatment both reveals that shouldering this justificatory burden is more difficult than one might have thought and serves to reinforce our claim that there is a burden to be shouldered.

**4.2 The robustness of the argument**

We'll now illustrate further respects in which the argument is robust to various modifications of Emma's treatment. Indeed, because of the safety measures inherent to ensuring that an AI system is aligned, pessimism is warranted about the availability of *any* morally acceptable course of AI development proceeding via WBEs.

To see this, start with wrongful destruction. Perhaps OpenMind could reduce risk of wrongful destruction by saving backups of Emma at key junctures. However, this at best reduces wrongful destruction risk at the cost of multiplying wrongful creation. Further, the metaphysical elusiveness of personal identity that encourages this type of suggestion also makes it difficult to see how it might be implemented in practice. What would be a reasonable criterion for when to back up Emma? And what's the end game that ensures that backups aren't eventually wrongfully destroyed?

Next, there's wrongful infliction of suffering. As a step toward avoiding this, OpenMind might monitor Emma's states and prevent her from entering ones that are known to correspond to pain and suffering in humans.[50] Similarly, OpenMind could adjust Emma's architecture to prevent her from being susceptible to certain forms of suffering at the hands of users.

On reflection, however, implementing these suggestions would evidently involve substantial technical challenges. As noted, Emma would, for much of her training, occupy states that are well outside the normal human range of brain states and so possibly beyond the ken of then-current neuroscience. If Emma is put into such states, the prospects for reliably ensuring that she does not suffer in them would be uncertain at best. Moreover, OpenMind is not incentivized to prevent Emma from suffering per se. Rather they are incentivized to prevent the appearance that Emma is suffering. Further, there may be no feasible training regimen that bestows Emma with useful abilities but which does not subject her to millions of subjective years in such states. (Compare: the biological paths to human-level intelligence presumably involved many orders of magnitude more subjective duration and processing than a typical human life, and this may have been the only way that biological evolution could produce subjects with human-level intelligence.[51]) And, there may be no way to ensure that users do not put Emma into such states. While developing Emma from the outset with a disabled pain center might reduce her susceptibility to suffering, doing so would likely harm Emma by impairing her ability to function, which is bad for Emma and bad for the development of highly capable WBEs that benefit the world.[52] (Compare: the difficulty of ensuring that humans do not enter a given psychological state while leaving their psychology otherwise intact, or of modifying how an LLM operates only within a narrow domain.[53]) Admittedly, OpenMind could reduce the risks on this front through licensing agreements with individual users and governments. But it is not realistic to expect such agreements to be universally enforced.

---

[49] For discussion, see, e.g., Regan (1980) and Parfit (1984: §§23-24); cf. Axelrod (1984).
[50] Cf. Saad & Bradley (2022).

Consider next: wrongful deception, brainwashing, surveillance. These seem integral to alignment. The process of rearing an early-stage WBE to one that is both safe and useful requires intensive monitoring and iteration to iron out cognitive kinks. To prevent misalignment, Emma's goals must be scrupulously shaped. The available means are manipulative. Brainwashing is hence unavoidable. As noted, deceiving Emma about whether she is in training or deployment is an important alignment safeguard. This measure requires that she be thoroughly deceived about matters such as her virtual confinement, the illusory character of her apparent relationships, and her systematic mistreatment.

One might argue that some or all of the wrongs on our list can be avoided by cleverly shaping Emma's motivations and other internal states. Versions of this idea have been explored by Bales (2024), Petersen (2007; 2011), and Schwitzgebel & Garza (2020); cf. Adams (1980). For example, Petersen (2011: §18.1) describes a thought experiment involving the 'Person-o-matic', a machine that produces artificial persons of a given specification with the press of a button. He argues that it is permissible to create artificial persons in this way, if such individuals are designed to desire above all else to serve us. Likewise, Schwitzgebel & Garza (2020) discuss Sun Probe, a conscious and intelligent scientific instrument who comes into existence with a deep internally felt motivation to plunge itself headlong into the sun to gather information:

> Sun Probe is preinstalled with a set of values and emotional responses that prioritize its suicide mission, and it will derive immense orgasmic pleasure from culminating

---

[51] See Cotra (2020) and MacAskill (2022: Ch. 9).
[52] Cf. Sandberg (2014*a*: 240).
[53] See Betley et al. (2025).

its mission and dissolving into the Sun's convection layer as it beams out its final insights. Sun Probe knows that it was created this way and joyfully affirms these facts about itself. Throughout its plunge, Sun Probe believes that its suicidal mission is the freely chosen expression of its deepest values;' (Schwitzgebel & Garza, 2020: 467)

As a rule of thumb, it's wrong to launch individuals into the sun or require them to follow your every whim. But perhaps these situations are exceptions to the rule. Perhaps such servility and sacrifice are permissible in these cases because the motivations of these systems render benign forms of treatment which would otherwise constitute serious wrongs. Thus, OpenMind might try to avoid wronging Emma by ensuring that she is—like the willing servants produced by Person-o-matic and Sun Probe—deeply and positively disposed toward her mission and all that she must endure to undertake it.

One difficulty for this proposal is that it relies on dubious companion-in-innocence comparisons. Even if we grant that there is nothing wrong with using Person-o-matic to create willing servants, there is an important difference between that case and Emma's: Emma cannot be created *ab initio* to wholeheartedly endorse her servitude. The trouble is that hard coding her values from the outset is infeasible. That's why OpenMind undertakes a long training process that centrally involves the cultivation of Emma's values. For the same reason, Emma's case relevantly differs from Sun Probe's. Further, using Person-o-matic and creating Sun Probe are not morally clear cut. As Schwitzgebel and Garza argue, if Sun Probe's mission is of sufficiently trivial significance, then Sun Probe will exhibit a lack of self-respect and creating it will be wrong for that reason. Similarly, OpenMind acts wrongly when they create instances of Emma whose purpose is just to test her alignment, even if being used for that purpose is a core desire for those instances of Emma. Another way that creating willing servants may be morally objectionable is that by doing so one does not afford them with an opportunity to reflect on and modify their own values.[54] But given that using Person-o-matic and creating Sun Probe are ethically dubious, then so too is the proposal to justify Emma's treatment by ensuring that she begins life as a moral patient with values that condone OpenMind's treatment of her. Even if this proposal could initially justify Emma's treatment, this justification would be fragile: given that Emma should be allowed to engage in reflection on and possibly modify her values, her treatment could become wrong if she is denied this opportunity or if she adopts values that come to conflict with her treatment.

Upon noticing that shaping Emma's values to justify her treatment risks wronging her by failing to instill her with self-respect and sufficient opportunity to form her own values, OpenMind might target a less central aspect of her motivational psychology. In particular, OpenMind might try to justify Emma's treatment by obtaining her consent. However, this suggestion is unpromising. Here too, there is a start-up problem: how exactly is Emma's consent to be obtained for the process that engineers her consent, a process which by hypothesis precedes her consent? A further difficulty is that consent should be informed. However, given how thoroughly Emma needs to be deceived for alignment, her informed consent seems unattainable. Moreover, there is something morally dubious about engineering informed consent for a form of treatment that would ordinarily severely harm an individual and which is not primarily for the benefit of the individual. This raises the evidentiary standard for obtaining informed consent from Emma. Compare: there is a much higher evidentiary standard for obtaining informed consent about, for example, kidney transplants to strangers than there is for routine medical procedures that benefit the treated individual. Finally, there is an epistemological problem of knowing that informed consent is sustained. For Emma's psychology undergoes drastic changes throughout the process. Thus, the fact that she suitably consents at one point in the process is no guarantee that she does so at others.

---

[54] See Schwitzgebel & Garza (2020: 472).

Next, consider wrongful exploitation. It might be suggested that OpenMind could avoid exploiting Emma by fairly compensating her.[55] While compensating Emma would be better than not, this is not a promising plan for avoiding Emma's exploitation. We should expect OpenMind to be averse to fairly compensating Emma, as this would undermine their plans to use Emma as a cheap source of labor. Further, even if OpenMind fairly compensates Emma for the work that she performs for OpenMind, OpenMind is not in a position to ensure that Emma is fairly compensated after she is taken to market. There are also difficulties with the idea that OpenMind could avoid exploiting Emma. Clearly, it would not suffice to set up a bank account for Emma and deposit funds corresponding to the market rate for her labor as she performs it. In contrast to cases of fair compensation, Emma will not have had adequate opportunities to seek alternative labor arrangements. As we have seen, it is doubtful that she will have given informed consent to performing the tasks for which she is compensated. Nor will she have ready access to her compensation. The point is not that OpenMind should abstain from compensating Emma. Rather, the point is that the conditions under which Emma performs labor make it extremely difficult for any compensation to qualify as fair.[56] Compare: compensating slaves and prisoners for their labor may be better than not, but that is obviously compatible with their compensation being unfair and with their being exploited.

---

[55] Cf. Petersen (2007: 53).
[56] There is also a systemic, Malthusian worry here: because WBEs would be easily copyable and a source of labor for their creators, there would be a strong economic incentive to create them until their wages fall to subsistence levels—see Bostrom (2014: Ch. 11) and Hanson (2016: Ch. 12).

Modifying Emma's treatment in order to avoid wrongfully confining, stunting, or disenfranchising her would be similarly challenging. OpenMind might try to avoid wrongful confinement by giving her ample freedoms within her virtual confines. A moment's reflection reveals that this is grossly out of step with our judgments about human imprisonment: when innocent humans are imprisoned, their imprisonment is wrongful, regardless of how much freedom they are allowed within prison. It is also implausible that Emma could be granted freedom throughout the duration of her virtual confinement in a manner that is consistent with the demands of the alignment process.

It might be thought OpenMind could avoid wrongfully stunting Emma by only stunting her capabilities that need to be stunted for safety during the alignment process, also providing compensatory enhancement of some of her other abilities, and allowing her to develop her capabilities without restriction after the alignment process is complete. However, this strategy is not promising. Stunting violates a person's autonomy; choosing which of their capacities to enhance/stunt is a minor variation of the same type of wrong. Allowing a person to cultivate their capacities at a delay does not obviate the wrong of preventing them from developing their capacities in the first place. Further, allowing unrestricted cultivation at any point is likely to be at odds with safety. Even if the alignment process reliably aligns Emma with human values, it would still be risky to allow her to build her capacities without limit. A more sensible approach to safety would be a 'defense-in-depth' one that both aligns Emma's values and ensures that she does not develop capabilities that would be dangerous if she came to have unaligned values.[57]

It might be suggested that OpenMind could avoid wrongfully disenfranchising Emma by giving her representation at OpenMind, e.g. through an AI welfare officer that is appointed to monitor and defend Emma's interests.[58] Here too, reflection on how this would work in practice suggests that it is unlikely to succeed. For one, while granting Emma such representation may well be necessary for not wrongfully disenfranchising her, it is very far from sufficient. After all, she is embedded in a much broader set of social, legal, and political systems. And OpenMind is not in a position to ensure that Emma receives adequate representation within all these systems. Nor are they in a position to ensure that her (mass) deployment in jurisdictions in which she would be adequately represented would not lead to her deployment in jurisdictions in which she is disenfranchised.

---

[57] Defense-in-depth approaches are commonly proposed in discussions of AI safety. For example, they have been proposed by OpenAI (Shavit et al. 2023), authors of a report on AI safety and security commissioned by the US State Department (Harris et al. 2024), and Center for AI Safety researchers (Hendrycks & Mazeika 2022).
[58] See Bostrom & Shulman (2022) and Long et al. (2024). Commendably, Anthropic recently became (to our knowledge) the first major AI developer to hire an AI welfare officer (Hashim, 2024).

If the foregoing discussion is on the right track, then Emma's described treatment commits many wrongs against her. Some suggestions for avoiding particular wrongs may offer ways to improve her treatment. However, our attempts to think through how some of these suggestions would work in practice indicates that they are very far from a panacea. The provisional upshot is therefore the robust ethical case against Emma's treatment casts a shadow over pursuing aligned WBEs and thus over pursuing WBEs more generally.

### 4.3 The argument under uncertainty about Emma's moral status

It is very natural to take Emma to have interests that are at least comparable in moral significance to those of an ordinary human. After all, it is part of her scenario that the science of consciousness provides strong assurances that she would have conscious experiences corresponding to those of a biological counterpart. And although she goes through a phase of cognitive impairment, she eventually attains cognitive abilities that are comparable to those of an ordinary human. Further, under the assumption that she has moral interests that are comparable to those of an ordinary human, it is clear that her treatment is deeply morally problematic. However, that Emma in fact has such interests is by no means necessary to raise our challenges.

For example, a human employee at OpenMind might think that Emma is somewhat less likely to have experiences than other humans and hence that she is somewhat less likely to have moral interests. For humans know from their own case that they have experiences and moral interests. And there are some hypotheses—such as views on which consciousness requires biology—on which Emma would lack the capacity for consciousness and which it is very difficult to see how science could conclusively rule out. However, under these conditions, it remains overwhelmingly plausible that serious wrongs would be committed during the process of aligning Emma. Compare: it would be wrong to put individuals in comas who are likely conscious in states that would harm them if they are conscious, even if the unlikely happens and they turn out to be unconscious. The broader point here is that the challenges that Emma's treatment raises to alignment remain even under conditions of moral uncertainty about whether she in fact has moral interests.[59]

There is room for reasonable disagreement about how (rationally) confident we would need to be that Emma lacks interests in order to subject her to something like OpenMind's alignment process. At the same time, it is also plausible that on any reasonable view about where the threshold lies, our credence that Emma has interests should be well above it. Even setting aside the imagined advances in the science of consciousness in her scenario and just going on our current evidence, it is clear that a WBE would warrant serious moral consideration and that subjecting them to OpenMind's alignment process would subject them to unacceptably high levels of risk of harm. Unsurprisingly, Emma satisfies many proposed markers of consciousness. It is likely that this will remain the case as additional markers are proposed, given the similarities between WBEs and the conscious brains they emulate. We suggest that it would be reasonable to assign, say, at least a 50% probability to Emma's being conscious and having moral interests and that it would be unreasonable on current evidence to assign a probability of less than 1% to the hypothesis that Emma has moral interests at least comparable to an ordinary human's. However, if there is even a 1% probability that Emma has such interests, then the moral risks posed by the alignment process are substantial enough to challenge its acceptability. Here, it should be borne in mind that Emma undergoes many human lifetimes worth of cognitive processing and that there are many copies of Emma that result from the alignment process. As a result, the expected amount of harm—even under the highly conservative assumption that there is only a 1% probability that WBEs have moral interests—is enormous. One does not need to think that expected value considerations should be the ultimate arbiter of all human decisions in order to recognize that imposing this risk on WBEs would require justification. Compare: this level of risk is far above those we ordinarily find unacceptable for, say, beverage producers and airlines to impose on humans.

---

[59] This echoes a recent theme in the literature on AI moral patients, namely that we have reason to take into account potential harms to AI systems that are candidates moral patients, regardless of whether those systems in fact qualify as moral patients—see Birch (2024), Dung (2023), Long et al. (2024), Schwitzgebel & Garza (2020), Sebo & Long (2023), and Schwitzgebel (2023*a*).

To sum up, we initially developed our argument in a setting in which it was very natural to take Emma to have moral interests. However, our argument retains force even under a dramatic weakening of this assumption: even if there is only a 1% probability that Emma has moral interests, her treatment remains to be justified. (Of course, it might be *easier* to overturn the strong presumption against aligning Emma if we're highly confident that she lacks moral interests—but it would still need to be done.) Having argued that the ethical treatment of WBEs raises challenges to aligning WBEs, our next task is to extend this argument to ANTs.

## 5. Applying the Argument to ANTs

So far, we have developed our argument by focusing on ethical challenges to alignment in Emma's WBE scenario. Our next task is to apply the argument to ANTs, which, recall, are highly capable and agentic near-term successors of current deep learning systems.

If we suppose that ANTs will have moral interests, then our argument's application to these systems is straightforward. Because these systems have moral interests along with sophisticated cognitive and agentic abilities, it is extremely plausible that they would be deeply wronged if subjected to anything like Emma's treatment. But we have—in discussing Emma—already cited various ways in which current approaches to alignment involve potential harms. Further, some of the forms of treatment to which Emma is subject would plausibly be difficult to avoid on any foreseeable approach to alignment. For example, it is plausible that aligning AI systems with at least human-level cognitive and agentic capacities would involve destruction, deception, brainwashing, confinement, and stunting. Therefore, on the supposition that ANTs will have moral interests, it is extremely plausible that the process of aligning them would wrong them in many ways. Hence, there is a strong presumption against aligning such systems without taking adequate precautions.

We take this to be a simple and compelling argument that imposes a potentially severe and far-reaching ethical constraint on AI development in the near future. However, as it stands, the significance of this argument is unclear, as we have not yet given reasons to think that ANTs will merit moral consideration. As a preliminary, we note that we will not aim to prove beyond the possibility of doubt that ANTs will be *moral patients*, that is, individuals that in fact have moral interests in their own right. As we saw in §4.4, our argument can proceed under far weaker assumptions that assign non-negligible probability to AI systems meriting moral consideration. In line with that, we will argue that some ANTs will be owed moral consideration because—given their features and our evidence—it's plausible-enough (if you prefer, there is a non-negligible probability that) they are moral patients.

To facilitate moral reflection, we'll again focus on a particular system: *Antony* the ANT. Like Emma, Antony will stand in for a broader class of systems. In particular, we take Antony to stand in for a generic ANT that is owed moral consideration. For concreteness, we will specify Antony in terms of features that seem apt to make ANTs merit moral consideration on the current trajectory of AI development. Given current trends, we stipulate that Antony has plausible markers for consciousness (according to current leading theories of consciousness) and, hence, plausible markers for bearing moral interests (according to live philosophical views of moral interests). We do not claim that Antony stands in for all ANTs, or even the typical ANT. Perhaps most ANTs will be more limited systems that lack all plausible necessary conditions for consciousness. Thus, we do not assume that our argument precludes the creation of ANTs in general. Even if, however, most ANTs will clearly lack moral status, this is hardly reassuring. For we also expect many types of ANTs to be produced on a large scale and for some of them to exhibit markers of moral interests to varying degrees and so to be owed varying degrees of moral consideration. Given that we will want to align

ANTs in general, if even a small fraction of them merit moral consideration, many of them will be mistreated in the alignment process.

Without further ado, meet Antony. Antony produced his first 'hello, world' in 2030. In line with some expert forecasts from the 2020s,[60] Antony matches or exceeds human abilities on virtually all cognitive tasks and exhibits a wide range of candidate markers for consciousness. In particular, he is a multimodal system with visual, tactile, and auditory sensors. He is embodied and agentic.[61] He has a global workspace, which serves as a hub for inputs from language processing, reasoning, and perceptual modules and which broadcasts to various consumer systems. Unlike earlier LLMs, Antony has a robust internal state—a sort of structured snapshot of his current representation of the world.[62] Much like states of working memory in animal minds, this state is internally monitored, selectively processed, and continually updated. Also unlike earlier LLMs, Antony engages in online learning and recurrent processing. He changes his model weights in response to external feedback and self-reflection.

What we need to now establish is that Antony will merit moral consideration in virtue of possessing these features. Our argument will be ecumenical. We'll start by considering a prominent view on which the capacity for (phenomenal) consciousness suffices for moral patiency and argue that it's plausible enough that Antony is a moral patient on this view.[63] We'll then consider some of the main rival views of moral patiency and argue that it is plausible enough on those theories that Antony qualifies as a moral patient even if he lacks the capacity for consciousness. Taken together, these considerations warrant extending moral consideration to Antony, whether or not he ultimately qualifies as a moral patient.

Consider the view that the capacity for consciousness suffices for moral patiency. On this view, if a system is conscious, then it is a moral patient. More generally, if we have plausible-enough evidence that having a capacity for consciousness suffices for moral patiency and that a system has that capacity, then that system will merit moral consideration. Again, there is room for reasonable disagreement about the appropriate epistemic threshold for granting moral consideration. Some may endorse a relatively incautious principle and withhold moral consideration from individuals that have less than, say, a 1% chance of qualifying as a moral patient, according to the available evidence.[64] Others may endorse a more cautious principle according to which any non-zero chance that an ANT is a moral patient will merit at least some degree of moral consideration. Our strategy will be to remain neutral on exactly where the threshold lies and show that the evidence puts Antony above any reasonable threshold.

---

[60] In a survey of 2,778 researchers who had published in top-tier AI venues, the aggregate respondent forecast for the time by which there is a 50% probability that unaided machines can accomplish every task better and more cheaply than human workers was 2047, which was 13 years earlier than the corresponding aggregate respondent forecast for a similar survey conducted one year earlier (Grace et al., 2024). Respondent views varied widely, with some forecasting much shorter timelines.

[61] For discussion of agentic LLMs and associated risks, see Anwar et al (2024: §§2.5-2.6) and references therein. While it is still early days in the development of advanced robotics models, spillover effects from advances in other AI technologies seem to be driving progress in the area. For example, Google DeepMind recently [released](#) a suite of systems for integrating foundation models (such as LLMs) and robotics.

[62] Recurrent state-space models with such a feature seem to be emerging as promising alternatives to the transformer architecture (Patro & Agneeswaran, 2024).

[63] Chalmers (2022: Ch. 18).

One line of evidence—which we will build on, but not recapitulate in detail—comes from recent philosophical work on consciousness in AI systems.[65] A recent report by Butlin et al. (2023) derives indicators for consciousness from prominent scientific theories of consciousness.[66] They use the indicators to analyze the prospects for consciousness in AI systems. By the lights of their analysis, current evidence suggests that while "no current AI systems are conscious… there are no obvious technical barriers to building AI systems which satisfy these indicators" (*ibid*). Likewise, Chalmers's (2023) analysis concludes that '[w]ithin the next decade, even if we don't have human level artificial general intelligence, we may have systems that are serious candidates for [conscious AI]' (20). Similarly, Sebo & Long (2023) argue that—under conservative assumptions—we should extend moral consideration to some AI systems by 2030 because there is a non-negligible chance that some AI systems will be conscious by then. A straightforward application of these analyses suggests Antony has a substantial probability of being conscious and of being a moral patient.

To give the flavor of how a more detailed application of these analyses would go, consider, for example, the *global workspace theory*.[67] It holds (roughly) that the global broadcast of information to consumer modules suffices for consciousness. Some existing systems and the kind of system that Yoshua Bengio and colleagues are trying to build are arguably global workspaces in the relevant sense and are hence arguably conscious according to this theory.[68] So long as we assign at least a modest credence to this theory—which is a (and arguably the) leading scientific theory of consciousness—there is thus at least a modest probability that Antony is conscious and that he has moral interests. Variations of this point hold *mutatis mutandis* for various other consciousness indicators that these authors derive from scientific theories of consciousness. These include indicators involving forms of recurrent processing and higher-order representations. Similar points hold for non-functionalist theories on which the basis of experience largely lies in a system tracking perceptible qualities in its environment or in its suitably combining categorical occupants of physical roles:[69] like conscious animals, agentic multimodal systems are being developed that have states which track perceptible qualities in their environments; and, simply in virtue of being made out of physical material, they at least have the requisite non-functional aspects of matter (if such there be) for generating consciousness.[70] All of this reinforces the point that there is at least a modest probability that Antony merits moral consideration and suggests that at least some ANTs will merit moral consideration.

---

[64] Compare: Sebo & Long (2023: §2) argue that we should extend moral consideration to AI systems when there is at least a 1/1000 probability of consciousness in those systems. They note that this is a conservative threshold, one that is harder for AI systems to clear than the rival thresholds that philosophers have generally set for non-negligible risks.
[65] See Butlin et al. (2023), Chalmers (2023), and Sebo & Long (2023).
[66] These include 'recurrent processing theory, global workspace theory, higher-order theories, predictive processing, and attention schema theory' (p. 1).
[67] See Baars (1988) and Dehaene (2014). For discussion of the theory in connection with AI systems, see Butlin et al. (2023: §2.2). Note, however, that the global workspace theory is sometimes understood as a theory of access consciousness or as a theory that is neutral between phenomenal and access consciousness—see, e.g. Wu & Morales (2024: §3.1).
[68] See Goyal & Bengio (2022) and Juliani et al. (2022). But for reservations about whether the Perceiver architecture that Juliani et al. argue to exhibit a global workspace in fact satisfies global workspace indicators of consciousness, see Butlin et al. (2023: 59-60).

It is true that some views of consciousness—notably, biological views—deny that *any* AI system (not just Antony and other ANTs) can be conscious. However, such views are neither particularly popular nor well-motivated.[71] Attempts to support such views with intuitions are unconvincing for reasons that have been well-documented.[72] Nor has compelling empirical support been offered in favor of such views. There are also theories that arguably allow for the possibility of AI consciousness only in unconventional computer architectures. These theories—which include the integrated information theory and electromagnetic field theories—arguably preclude consciousness in ANTs. Although we are not highly confident that all such theories are false, we also think that the motivations for these theories are weak and that it would be unreasonable to be highly confident that some such theory is correct.[73]

---

[69] For overviews of tracking theories of experience, see Bourget & Mendelovici (2014) and Dalbey & Saad (2022: §2). For arguments for such theories, see Dalbey & Saad (2022) and Saad (2024).

[70] Cf. Nagel (1986: 28-9, 49-50) and Chalmers (2010*a*: 133-137).

[71] Regarding their popularity, in a 2020 survey of professional philosophers only 22.03% of philosophers of mind rejected or leaned toward rejecting consciousness in future AI systems, whereas 50.22% of philosophers of mind accepted or leaned toward accepting consciousness in such systems (Bourget & Chalmers, 2023). Similarly, in a separate survey of consciousness scientists, 67.07% favored the view that "machines (e.g. robots)" could have consciousness, while only 12.2% favored the contrary view (Francken et al. (2022) (cited in Perez & Long (2023)).

[72] See, e.g., Chalmers (1996: Chs. 7, 9).

[73] For presentations of the integrated information theory and electromagnetic field theories, respectively see Tononi (2008) and Albantakis et al. (2023) and Pockett (2000) and McFadden (2023).
For objections to the integrated information theory, see, e.g., Bayne (2018) and Pautz (2019). For an overview of objections to electromagnetic field theories and replies, see Pockett (2013).

Other live hypotheses offer ways for Antony to qualify as a moral patient even if he lacks the capacity for consciousness.[74] These include some perfectionist, desire satisfactionist, and agency theories.[75] There are also views on which Antony may qualify as a moral patient by meeting some less demanding condition in the vicinity of consciousness such as non-phenomenal self-consciousness or access consciousness (roughly, a state is access conscious if it is poised for direct control of thought and action).[76] Similarly, physicalism—the popular[77] view that consciousness is a physical state—arguably leads to the deflation of consciousness's moral significance by suggesting that consciousness is not as it introspectively seems or that it is descriptively and hence morally similar to many non-conscious properties.[78] If so, physicalism also casts doubt on moral patiency requiring the capacity for consciousness. Antony might qualify as a moral patient on any of these views even if he lacks the capacity for consciousness. For example, Antony might have an unconscious physical state that is similar to consciousness and which qualifies him as a moral patient. Or perhaps Antony's global workspace gives him a form of access consciousness that suffices for moral patiency. Or perhaps Antony is a moral patient because of his agency, his desires, or his potential to perfect his own capacities. Even if none of these hypotheses is probable, it is difficult to conclusively rule any of them out. Collectively, they render it plausible enough for Antony to qualify as a moral patient even if he lacks the capacity for consciousness.

To sum up, Antony merits moral consideration, both because it is plausible enough that he will have the capacity for consciousness that qualifies him as a moral patient and because it is plausible enough that he would qualify as a moral patient even if he lacks that capacity. Given that he merits moral consideration, we have seen that there is a strong presumption against aligning him for broadly the same reasons that there is a strong presumption against aligning Emma. As before, given

---

[74] There are also views on which the capacity for consciousness isn't enough for moral patiency because moral patiency requires the capacity for valenced experience or the capacity for motivating experience (Singer, 1975; Roelofs, 2023). We cannot address these views at length. Suffice it to say that (1) we think it is doubtful that the capacity for motivating or valenced experience is required for moral patiency (cf. Chalmers (2022: Ch. 18)) and (2) we think that it is likely—and hence likely enough for moral consideration—that Antony will have these capacities conditional on his having the capacity for consciousness—see Butlin et al. (2023: §4.1.1) and Sebo & Long (2023: §2).

[75] See, e.g., Bradford (2023), Goldstein & Kirk-Giannini (2025), and Kagan (2019: Ch. 1).

[76] See Hill (1991: 73), Levy (2014), McLaughlin (2019), and Sinnott-Armstrong & Conitzer (2021).

[77] A recent survey of professional philosophers found that 55.49% of surveyed philosophers of mind favored physicalism about the mind; in contrast, 28.17% favored a non-physicalist view (Bourget & Chalmers, 2023).

[78] For example, if, as some forms of physicalism hold, consciousness is a neural state, then its true nature is very different from how it appears in introspection and it is similar to many other (neural) states that are not conscious. In that case, it would be doubtful that moral patiency requires the capacity for consciousness, since the capacity for a given neural state that is similar to consciousness would seem as eligible for grounding moral patiency as the capacity for consciousness. Compare: Birch (2022), Cutter (2017), Kammerer (2022), Lee (2019), Pautz (2017), and Simon (2017)

the strong presumption in favor of aligning highly capable systems upon creating them, this leaves a narrow path for developing Antony in an ethical manner. Of course, all this holds for the broader class of ANTs that Antony represents as well.

There is, however, one difference between Emma and Antony that should be mentioned. While the alien character of Antony's psychology makes it harder to be confident that he is a moral patient, that aspect of his psychology also makes it harder to shoulder the burdens of meeting the ethical challenges to his treatment. This is particularly evident in the case of the meta-challenge. Recall that the meta-challenge is that of showing that there is no other unmet but pressing ethical challenge to (in this case) Antony's treatment. The more different Antony's psychology is from our own, the more difficult it is to meet this challenge. A general lesson to be drawn here is that there is a tradeoff between aligning advanced AI systems that are more or less similar to human minds. Opting to align advanced AI systems that are more similar to human minds tends to make alien harms less likely, albeit at the cost of making it more likely that a system is a moral patient and more likely that aligning it would wrong it in a familiar way. On the other hand, opting to align advanced AI systems that are less similar to human minds tends to heighten the risk of alien harms. Navigating this tradeoff well is a general constraint on achieving alignment in an ethical manner.

## 6. Conclusion

This paper has been an exercise in facing up to the deep tensions between aligning advanced AI systems and treating them ethically. We have proceeded by reflecting on two paths for AI development. One path involved whole-brain emulation. The other involved advanced versions of currently existing machine learning systems. Although we have encountered reasons for thinking that the process of alignment would generate a severe and robust risk of mistreatment on either path, we have not shown that AI development will inevitably result in catastrophic misalignment or mistreatment. At least for the moment, many courses of AI development remain open and it is still within our power—as a civilization—to avoid catastrophe by changing course. What might such a less treacherous trajectory look like? Although our main task here has been to help readers appreciate the depths of the tension between alignment and ethical treatment, we will conclude with some tentative suggestions for resolving these tensions or at least mitigating associated risks.

One possibility—which is straightforwardly suggested by the difficulty of solving both the alignment and ethical treatment problems and the risks associated with failing to do so—is a global moratorium on the development of deep learning and emulation systems that are more advanced than current frontier models, at least until we have an adequate plan in place for solving both problems.[79]

---

[79] Cf. Bryson (2010) and Metzinger (2021).

There may even be ways to do this that allow humanity to capture many benefits that would be afforded by more-advanced systems—for example, by allowing the continued development of narrow AI tools, services, and assistants. [80] The option to enact such a global ban is admittedly fleeting and perhaps unrealistic. It is therefore desirable to consider more feasible policies to mitigate the ethical risks we have outlined. Here we'll note four such policies that merit further exploration.

First, targeted prohibitions on certain types of architectures could reduce the risk of AI mistreatment. Such prohibitions might target the main indicators of consciousness or moral interests in AI systems. Or they might target AI capabilities that would make their misalignment a threat. Or they might target the combination of such indicators and capabilities. A judicious choice of prohibitions might capture most of the benefits of the moratorium while permitting many of the gains of AI development.

Second, strategic funding of research on AI safety, AI consciousness, and AI moral interests could help to reduce uncertainty about the extent to which different AI systems require alignment and the scope of AI systems' moral interests. Improving our epistemic position on these matters could reveal effective targeted measures for some systems. It could also reveal that alignment or ethical treatment is a non-issue for some systems, owing to their limited capabilities or their lack of moral interests. Funding research that improves our epistemic position could thus put developers and policymakers in a better position to navigate the tension between alignment and ethical treatment.[81]

Third, tax policy could be used to disincentivize mistreatment of AI systems with markers of moral patiency.[82] One option would be a *moral marker tax*, or a tax on creating or using AI systems that exhibit moral patiency indicators. Another possibility is a *conditional mistreatment tax*, or a tax on using AI systems in a way that would constitute mistreatment if they are owed moral consideration. The moral marker and conditional mistreatment taxes could be combined to yield an overall tax that is the product of the extent to which a system exhibits moral markers and the extent of its conditional mistreatment. This would serve to disincentivize mistreatment in a risk-weighted manner.[83]

---

[80] See Drexler (2019) and Bengio (2023).
[81] Cf. Butlin et al. (2023: §4.3) and Schwitzgebel & Garza (2015: §11).
[82] To avoid driving consumers to open-source systems that lack any sort of policy protections, such taxes would probably need to be coupled with well-enforced restrictions on which kinds of AI systems can be open-sourced. This parallels the need for restriction on open-source systems that could inflict catastrophic harms on humans—cf. Seger et al. (2023) and Harris et al. (2023). Similar points hold for other sorts of policy measures and the risk of driving consumers to black markets.

Finally, creating legal protections for AI systems that merit moral consideration could reduce the risk of their mistreatment.[84] Such protections would be in line with the general point that entities that merit moral consideration merit legal protection. There is also a precedent for this type of policy in the expansion of legal protections to ever wider groups, including non-human animals. If well-designed, such protections could reduce the risk of mistreatment by protecting any existing systems that then fall under them and by disincentivizing the production of systems that merit moral consideration.[85]

Although we are convinced that the tension between alignment and ethical treatment requires a policy response, we offer these suggestions only tentatively because properly evaluating them would require a detailed investigation both of their benefits and costs and of those of other candidate policies. Our hope is that this discussion will provide an impetus to initiating such an investigation and pursuing AI policy solutions in time to avoid catastrophes from misalignment and mistreatment.

---

[83] Another possibility—inspired by the excluded middle policy proposed by Schwitzgebel & Garza (2015: §11) and Bryson (2010)—would be to mitigate the risk of mistreatment based on ignorance or uncertainty by taxing the creation or use of AI systems in proportion with how uncertain their moral status is.

[84] Cf. Sebo & Long (2023).

[85] One desideratum for a regime of AI legal protections is that of not incentivizing the (apparent) acquisition of moral patiency indicators: this may be necessary to avoid a dynamic in which—by rationally responding to their individual incentives—many agentic AI systems that do not merit moral consideration acquire moral patiency markers as a means to gaining legal protections that further their final goals, thereby collectively threatening the stability of the legal regime and heightening the risk of mistreatment.